# Absolute frequency measurement of $^1S_0$ (F = 1/2) – $^3P_0$ (F = 1/2) transition of $^{171}$Yb atoms in a one-dimensional optical lattice at KRISS

Chang Yong Park, Dai-Hyuk Yu*, Won-Kyu Lee, Sang Eon Park, Eok Bong Kim, Sun Kyung Lee, Jun Woo Cho, Tai Hyun Yoon**, Jongchul Mun, Sung Jong Park, Taeg Yong Kwon, and Sang-Bum Lee

Korea Research Institute of Standards and Science, Daejeon 305-340, Korea
*Corresponding author e-mail: dhyu@kriss.re.kr
**Current address: Department of Physics, Korea University, Seoul 136-713, Korea

We measured the absolute frequency of the optical clock transition $^1S_0$ (F = 1/2) – $^3P_0$ (F = 1/2) of $^{171}$Yb atoms confined in a one-dimensional optical lattice and it was determined to be 518 295 836 590 863.5(8.1) Hz. The frequency was measured against Terrestrial Time (TT; the SI second on the geoid) by using an optical frequency comb of which the frequency was phase-locked to an H-maser as a flywheel oscillator traceable to TT. The magic wavelength was also measured as 394 798.48(79) GHz. The results are in good agreement with two previous measurements of other institutes within the specified uncertainty of this work.

## 1. Introduction

Recently optical clocks based on the narrow optical transitions of neutral atoms in an optical lattice [1-18] or a single ion in a Paul trap [19-21] have proved their potential capability to reach uncertainty levels below $10^{-17}$, and some of them have already surpassed the best microwave clocks in terms of stability and systematic uncertainty [22-24]. Compared to optical ion clocks, optical lattice clocks are expected to have much better short-term stability with a large number of atoms interrogated, which may lead to much smaller quantum projection noise. Sr [1-6], Yb [7-14], Hg [15-17], and Mg [18] atoms are being actively investigated for future optical lattice clocks in many institutes worldwide.

Among these elements, Yb has several merits for an optical lattice clock. Natural abundance of Yb isotopes is reasonably distributed over bosonic and fermionic isotopes. Adequate transitions exist for manipulating atoms in an ultracold state. Furthermore, solid-state lasers are available for all related processes to investigate the clock transition [25-27]. Ytterbium optical lattice clocks using a bosonic isotope ($^{174}$Yb) were first realized with the magnetically-induced spectroscopy method [7-10]. Fermionic $^{171}$Yb atoms have also been actively studied with the merit of reduced collision shift due to the Pauli exclusion principle. Additionally, the simplest spin-1/2 system has been found to provide effective single-stepped optical pumping between Zeeman sub-levels for preparing a spin-polarized ensemble of atoms, and the frequency shift associated with the tensor polarizability induced by the lattice electric field is intrinsically removed.

Currently, at least five groups are studying the use of $^{171}$Yb for an optical lattice clock, to our knowledge, and two groups have reported measurement results on the absolute frequency of the clock transition [12, 13]. Frequency comparison between laboratories plays a crucial role in the discussion of the next-generation frequency standard, since independent measurements in many laboratories are required to confirm that there are no remaining systematic errors as in the case of the Sr optical lattice clock [1-6]. In this paper we report the third independent absolute frequency measurement result, to our knowledge, of the optical clock transition of $^{171}$Yb atoms in a one dimensional (1-D) optical lattice. Frequency shifts due to the optical lattice, Zeeman effect, blackbody radiation, collisions between trapped atoms, and gravitational effect were investigated to evaluate the shifts and systematic uncertainties. The absolute frequency was measured using an optical frequency comb, which was referenced to an H-maser as a flywheel oscillator. With the final correction of the H-maser frequency offset against Terrestrial Time (TT; the SI second on geoid), the absolute frequency of the transition was determined to be 518 295 836 590 863.5 Hz, with an uncertainty of 8.1 Hz ($1.5\times10^{-14}$). This result agrees well with the two previous measurements performed by other laboratories within the uncertainty of this work.

In section 2, the whole experimental scheme and laser systems are overviewed. In section 3, the systematic shifts and the uncertainties of this measurement are discussed. In section 4, the absolute frequency is determined with the total uncertainty, including the systematic uncertainty, the statistical uncertainty, and the uncertainty related to the link to TT. In section 5, conclusions and future plans are given.

## 2. Experiment

The clock transition $^1S_0\,(F = 1/2) - {}^3P_0\,(F = 1/2)$ and other relevant energy levels of $^{171}$Yb atoms are depicted in Fig. 1. The strong transition of $^1S_0\,(F = 1/2) - {}^1P_1\,(F = 3/2)$ at 399 nm was used for Zeeman slower and first-stage magneto-optical trap (blue-MOT). Then, the transition of $^1S_0\,(F = 1/2) - {}^3P_1\,(F = 3/2)$ at 556 nm was used for the deep cooling (green-MOT) to prepare Yb atoms in a 1-D optical lattice providing the Lamb-Dicke condition. After the clock transition was interrogated with a clock laser, the normalized population of the excited state and the ground state was measured.

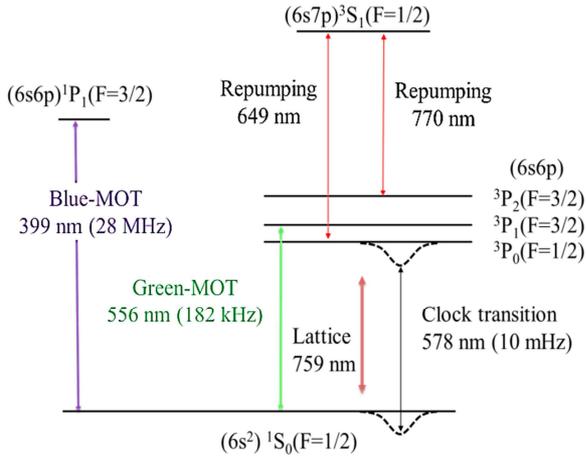

**Figure 1.** Relevant energy levels of $^{171}$Yb atoms and lasers to drive various transitions. Lasers for optical lattice (759 nm), clock transition (578 nm), two cooling transitions (399 nm and 556 nm), and two repumping transitions (649 mm and 770 nm). Natural linewidths of the cooling transitions and the clock transition are given in parentheses.

### 2.1 Laser systems

Fig. 2 shows a schematic diagram of the experimental setup. The laser system for the Zeeman slower and the blue-MOT consists of two Fabry-Pérot diode lasers (F-P LD's in Fig. 2) (NICHIA; NDV4313; 120 mW, 400~401 nm wavelength selected) injection-locked by a master external-cavity diode laser (ECDL). The slave lasers kept at 13°C were seeded with injection power of 1 mW. The frequency of the master ECDL was locked to the fluorescence signal of the $^1S_0(F=1/2) - {}^1P_1(F=3/2)$ transition obtained from an atomic beam. The frequency detuning of the blue-MOT beam was adjusted by variation of the intersection angle between the master laser and the collimated atomic beam for the best trap efficiency. The frequency of the seed laser for the Zeeman slower was shifted by -500 MHz from the resonance of the transition through a double-pass acousto-optic modulator (AOM). Both slave lasers had the output power of 40 mW, which was at 70~80% of the full power level to increase the lifetimes of the laser diodes.

The 556-nm light for the green-MOT was obtained by second harmonic generation (SHG). The output power of 20 mW from an ECDL (Toptica; DL100) at 1112 nm was amplified by a commercial Yb-doped fiber amplifier (YDFA, Keopsys; KPS-CUS-YFA-1111-SLM-10-PM-CO) up to 200 mW and converted to 556-nm light with an output power of 50 mW by using a ridge-type waveguided periodically-poled lithium niobate (WG-PPLN) crystal. As the natural linewidth of the transition for the green-MOT is $\gamma = 2\pi \times 182$ kHz, the linewidth of the ECDL was narrowed to about 10 kHz by the Pound-Drever-Hall technique [28] with an optical cavity (finesse 10,000). Then, the frequency of the 556-nm laser system was stabilized to the $^1S_0(F = 1/2) - {}^3P_1(F = 3/2)$ transition by side-of-fringe locking to a fluorescence signal from the atomic beam. The error signal was fed-back to the piezoelectric transducer (PZT) attached to a mirror of the optical cavity. The linewidth of the fluorescence signal was about 6 MHz due to the residual Doppler broadening. However, with the help of the laser power stabilization, the frequency drift during measurement was below 50 kHz.

The lattice laser was obtained from a commercial Ti:sapphire laser system (Coherent Inc.; MBR-110) pumped by a single-mode, 532-nm laser (Coherent Inc.; Verdi-V18), producing output power up to 3 W at 759 nm. Its frequency was stabilized to an internal reference cavity and was continuously measured using a calibrated commercial wavelength-meter with an accuracy of 10 MHz. After passing through an optical isolator, an AOM for power stabilization, and a single-mode optical fiber for mode cleaning, a vertical 1-D optical lattice was formed with laser power up to 0.8 W at 759 nm.

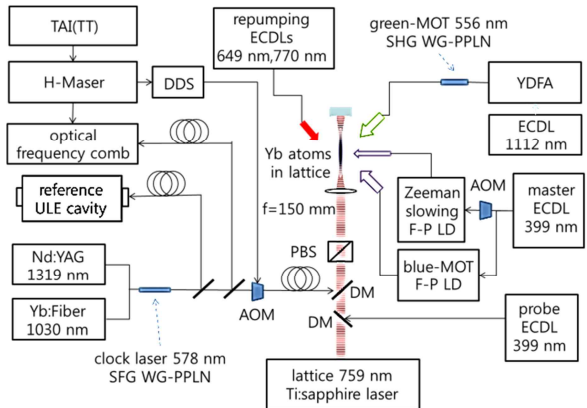

**Figure 2.** Schematic diagram of the experimental setup. 759-nm (lattice), 578-nm (clock), and 399-nm (probe) lasers are combined with two dichroic mirrors after transfer by optical fibers. TAI; International Atomic Time, TT; Terrestrial Time, DDS; direct digital synthesizer, SFG; sum-frequency generation, SHG; second harmonic generation, WG-PPLN; waveguided-periodically-poled lithium niobate, AOM; acousto-optic modulator, ECDL; external-cavity diode laser, F-D LD; Fabry- Pérot laser diode, YDFA; ytterbium-doped fiber amplifier, DM; dichroic mirror, PBS; polarizing beam splitter.

The clock laser at 578 nm was obtained from the sum-frequency generation (SFG) of a 1030-nm fiber laser with a 20 kHz linewidth (Koheras; ADJUSTIK) and a 1319-nm Nd:YAG laser with a 1 kHz linewidth (Innolight; Mephisto). A WG-PPLN was pumped simultaneously by the co-propagating 1030-nm laser of 10 mW and the 1319-nm laser of 20 mW producing a collimated output beam of 2 mW at 578 nm (Fig. 2). The SFG output was transferred by using polarization-maintaining single-mode optical fibers to three parts, namely, a super-cavity made of ULE glass for linewidth narrowing, an optical lattice setup for clock spectroscopy, and an optical frequency comb for frequency measurement. Frequency noises originating from the fiber transfers to the super-cavity and to the optical lattice setup were actively compensated [29]. The Pound-Drever-Hall technique was used to stabilize the clock laser frequency to a resonance of the super-cavity. A PZT of the 1319-nm Nd:YAG laser was used for the fast locking of the laser frequency on a cavity resonance, and a PZT of the 1030-nm fiber laser was used for the slow locking to compensate the slow drift. The super-cavity had a finesse of 350,000 and a free-spectral-range of 1.5 GHz. It was installed horizontally inside a vacuum chamber with the pressure of $10^{-5}$ Pa, and the temperature of the cavity was stabilized at around 30°C within 1 mK. The vacuum chamber was mounted on an active vibration isolation platform and the whole setup was enclosed by an anti-acoustic chamber with a 20-dB isolation.

To evaluate the frequency stability of the clock laser, we constructed an independent second clock laser system, which was based on the second harmonic generation of an 1156 nm laser [30]. The independent super-cavity in the second clock laser system was manufactured using the same ULE batch as in the clock laser system using SFG. Also, the cavity design was the same, including the finesse and the free-spectral-range. The second cavity was placed in a similar environmental condition as the first one. The beat note between the two independent clock lasers is shown in Fig. 3(a) with the frequency span of 500 kHz. The control loop bandwidth was evaluated to be about 30 kHz from the servo bump in Fig. 3(a), which was confirmed again also by the servo bump in the PDH error signal used in the frequency stabilization of the SFG system. The beat note between the two clock lasers with the frequency span of 5 kHz (resolution bandwidth: 47 Hz, sweep time: 43 ms) is shown in Fig. 3 (b). The black line is a typical single-sweep result, showing a full width at half maximum (FWHM) of 80 Hz. However, this FWHM linewidth varied mostly between 70 ~ 170 Hz in each sweep due to frequency jitters. The gray line is a result with maximum hold for 10 s, showing mid-term jitter of about 700 Hz in a measurement time of 10 s. The relatively large short-term linewidth and frequency jitter in our system was attributed to several causes. First of all, the cavity was not supported at the vibration-immune points [31], and the residual vibration level was relatively high even after an active vibration isolation platform was used, since the cavity setup was located on the second floor of the building. In addition, the control loop bandwidth (30 kHz) was insufficient to completely remove the clock laser frequency noise. Also, there was long-term frequency drift of the clock laser due to the room temperature variation, since the stabilized temperature was away from the temperature for the zero in the coefficient of thermal expansion (CTE) of the cavity. The temperature for a zero CTE was expected to be about 11°C according to the manufacturer's specification. However, we were not able to cool down the cavity to this temperature due to the difficulty in extracting the heat from the anti-acoustic chamber. However, the frequency drift was low enough to be calibrated by an optical frequency comb referenced to an H-maser. The typical drift rate of the clock laser was within ±4 Hz/s.

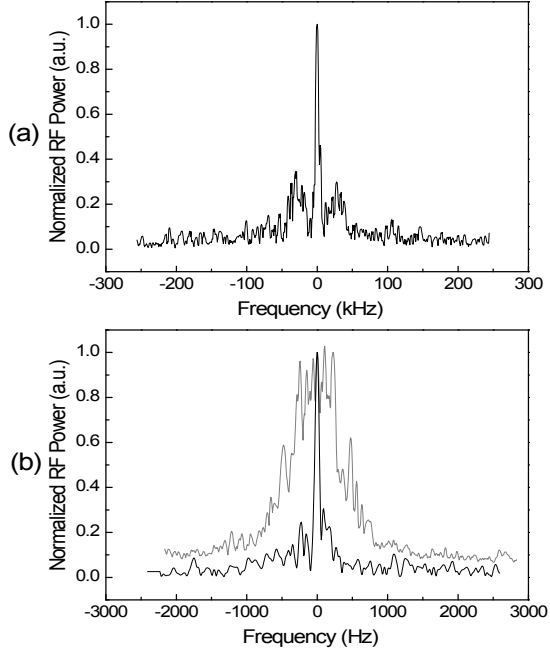

**Figure 3. (a)** Beat note between the two independent clock lasers (resolution bandwidth: 4.7 kHz; video bandwidth: 4.7 kHz; span: 500 kHz; sweep time: 27 ms). The control loop bandwidth is evaluated to be about 30 kHz. **(b)** Beat note between the two independent clock lasers (resolution bandwidth: 47 Hz; video bandwidth: 47 Hz; span: 5 kHz; sweep time: 43 ms). Black line; a typical single-sweep result, showing a FWHM linewidth of 80 Hz. Gray line; a result of maximum hold for 10 s, showing mid-term frequency jitter of about 700 Hz in a measurement time of 10 s.

Two ECDLs at 649 nm and 770 nm with the output power of 20 mW were built for the repumping of the populations in the two excited states, $^3P_0$ and $^3P_2$, to the ground state for the purpose of normalization of the number of atoms excited by the clock laser. The frequencies of the 649-nm and 770-nm ECDLs were stabilized by using the fluorescence signals from Yb atomic beam apparatus corresponding to the transitions of $^3P_0$ (F=1/2) – $^3S_1$ (F=1/2) and $^3P_2$ (F=3/2) – $^3S_1$ (F=1/2), respectively. A third ECDL at 680 nm was used to populate the $^3P_0$ and $^3P_2$ states in the atomic beam via the transition of $^3P_1$ (F=3/2) – $^3S_1$ (F=1/2) with the help of the frequency-stabilized 556-nm laser [32].

## 2.2 Experimental overview

The timing sequence for the experiment is shown in Fig. 4. A furnace containing pieces of natural Yb metal (purity 99.9 %) was heated to 450°C, and a collimated Yb atomic beam with seven isotopes was produced at a rate of ~$10^{10}$/s. The velocities of the $^{171}$Yb atoms were slowed in a 30-cm-long Zeeman slower. The slowing laser had a frequency detuning of -500 MHz from the resonant frequency of the $^1S_0$ (F = 1/2) – $^1P_1$ (F = 3/2) transition (natural linewidth; 28 MHz) with an intensity of 30 mW/cm$^2$. We estimated the flux of the slowed $^{171}$Yb atoms at the MOT region to be ~ $10^9$/s.

The same transition was used for the first-stage trapping of $^{171}$Yb atoms (blue-MOT). The laser for the blue-MOT had a power of 6 mW per axis with a 15-mm beam diameter. The magnetic field gradient by the MOT-coil was 3 mT/cm. The trapped atom number in the blue-MOT was saturated roughly at $10^8$, which was measured by a calibrated photo-multiplier tube (PMT). The number of trapped atoms did not increase further due to the shadow effect of cold atoms, which broke the intensity balance between the incident and the retro-reflected cooling lasers and made the atom cloud unstable. Even though the loading time constant of the blue-MOT was 410 ms, we used 180 ms to reduce the cycle time.

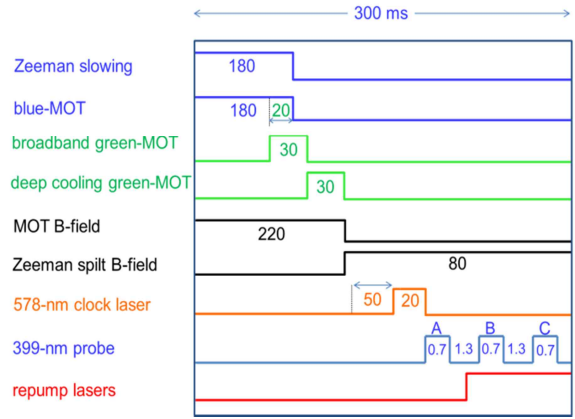

**Figure 4.** Timing sequence of the experiment for the interrogation of the clock transition of Yb atoms in an optical lattice (time unit; ms).

The Doppler temperature limit of the blue-MOT is about 0.7 mK, which is much higher than the potential depth of the optical lattice. Therefore, we utilized the second-stage trapping (green-MOT) by using the $^1S_0$ (F = 1/2) – $^3P_1$ (F = 3/2) intercombination transition at 556 nm with a narrow linewidth (182 kHz), and the corresponding Doppler temperature limit is 4 μK. This stage was divided into two steps as shown in Fig. 4. In the first step for 30 ms, a relatively high intensity (3 mW/cm$^2$ per axis) of the green-MOT laser, with the frequency detuned (-2 ~ -3 MHz) and modulated (2-MHz depth), was used to increase the atom number transferred from the blue-MOT to the green-MOT. At the end of this step, typically about 70% of the atoms trapped by the blue-MOT were transferred to the green-MOT with a temperature of 100 μK. In the second step for 30 ms, for the purpose of deep cooling, the power of the green-MOT laser was lowered to 0.5 mW per axis, the frequency modulation was turned off, and the center frequency was detuned to about -100 kHz from the resonance frequency. The gradient of the MOT magnetic field was maintained to be 3 mT/cm throughout the green-MOT stage.

After the green-MOT stage, the atoms were cooled down to about 30 μK, and the green-MOT laser was turned off using the AOM and a mechanical shutter. A linearly polarized lattice laser at 759 nm was aligned vertically and was always turned on throughout the measurement sequences. The laser was focused with an achromatic lens with a focal length of 150 mm and retro-reflected from a concave mirror. It has a dichroic coating for the 759-nm light with a high reflectance and for the 578-nm light with a high transmittance. The beam waist of the focused lattice laser was 25 μm, and the potential depth of the lattice was about 400 $E_r$, where $E_r$ is the recoil energy of the lattice laser with $E_r/k_B \approx$ 100 nK. In this way, roughly 1% of the atoms ($10^5$) could be transferred to the 1-D lattice from the green-MOT. The corresponding motional sideband frequencies of the longitudinal and radial motion in the lattice potential were estimated to be 50 kHz and 200 Hz, respectively. The lifetime of trapped atoms in the lattice was measured to be about 400 ms. The shot-to-shot number fluctuation of trapped atoms in the lattice was typically around 30%.

In the next stage, the clock transition, $^1S_0$ (F = 1/2) – $^3P_0$ (F = 1/2), was probed. The MOT-field was turned off by a MOSFET switch, and a small bias B-field of 430 μT was turned on by use of a pair of rectangular coils to split the two π-transitions. The spatial inhomogeneity of the B-field was estimated to be less than 1% in the region of the lattice trap, and the current fluctuation in the coil was also controlled within 1%. To avoid the ringing of the B-field immediately after the MOT-field was turned on, we waited 50 ms before introducing the clock laser. The direction of the bias B-field and the polarization of the clock laser allow only π-transitions among the Zeeman sub-levels. The clock laser was on for 20 ms with power of 100 nW. Longer interrogation time did not reduce the linewidth of the spectrum due to the broad short-term linewidth and frequency jitter of the clock laser. The clock laser was co-aligned with the lattice laser within 1-mrad accuracy, and the spot size at the interaction region was 90 μm so that it provided quite a uniform electric field for trapped atoms. The frequency of the clock laser was scanned typically with a 10-Hz step by a direct digital synthesizer (DDS), which drove a double-pass AOM.

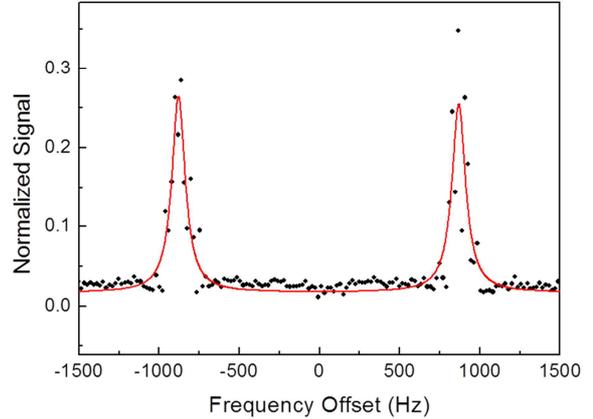

**Figure 5.** Typical normalized spectrum of the clock transition obtained by a single scan of the clock laser. The direction of bias B-field and the polarization of the clock laser allow only π-transitions. Red solid-line is the fit result of the spectrum with a double-peak-Lorentzian function.

The excitation probability of the clock transition was detected by measurement of the fluorescence signal from the atoms excited by a 399-nm ECDL resonant to the $^1S_0$ – $^1P_1$ transition [13]. Since the transition strength was sufficient, all the atoms in the ground state could be swept out within 0.7 ms, and the fluorescence signal was detected by a PMT. The fluorescence signal was integrated with a gated integrator for 0.4 ms (signal *A*). Excited state atoms were transferred to the ground state by the use of two repumping lasers at 649 nm and 770 nm resonant to the $^3P_0$ (F=1/2) – $^3S_1$ (F=1/2) and $^3P_2$ (F=3/2) – $^3S_1$ (F=1/2) transitions, respectively.

After the atoms were pumped to the ground state, the fluorescence signal was detected again (signal *B*). A third integrated signal (signal *C*) was used to eliminate the background noise. The dominant background noise was due to the scattered photons from the atomic beam. The normalized signal $S_N$ was calculated from the relation $S_N = \frac{b}{a+b}$, where $a = A - C$ and $b = B - C$. Fig. 5 shows a typical spectrum of the clock transition obtained by a single scan of the clock laser. About 25% of the atoms were excited at resonance, because we did not use the optical pumping between the two Zeeman sub-levels. The center frequencies of two π-transitions were separated by 1.8 kHz due to the applied bias B-field, and the typical linewidth of the spectrum of the clock transition was about 200 Hz due to the linewidth and frequency jitter of the clock laser as noted in section 2.1. To find the clock transition frequency, we fit the spectrum with a double-peak-Lorentzian function and obtained the average value of the two peak frequencies. The clock laser frequency was simultaneously measured by using an optical frequency comb referenced to an H-maser. Since the clock laser frequency was measured with the gate time of one second and the clock cycle time was 300 ms, we first fit the measured frequency data and used the result to estimate the clock laser frequency at an arbitrary clock interrogation time. The typical resulting uncertainty due to the clock laser frequency estimation by data fit was around 30 Hz, and this was the dominant uncertainty factor for each absolute frequency measurement. During the experiment, timing synchronization between the spectrum measurement and the frequency measurement was maintained within 10 ms. Since the drift rate of the clock laser was within ±4 Hz/s, a 10 ms timing difference can only result in a 40 mHz shift at most. Considering the variation of the clock laser drift rate (magnitude and sign), the actual shift was estimated to be much smaller than 40 mHz and was negligible for our experiment.

### 3. Evaluation of systematic frequency shifts

To experimentally evaluate some of the systematic shifts and uncertainties explained in this section, we measured the clock transition frequency shifts for a given experimental condition by using an optical frequency comb referenced to an H-maser. The H-maser had a relative frequency drift rate of $1.33(12) \times 10^{-16}$/day estimated from the Circular-T data. This drift could cause only 0.07 Hz shift for one day period. This drift was considered during the evaluation, but it was dominated by other uncertainties, mostly from the measurement uncertainty of the clock laser frequency by fit as noted above.

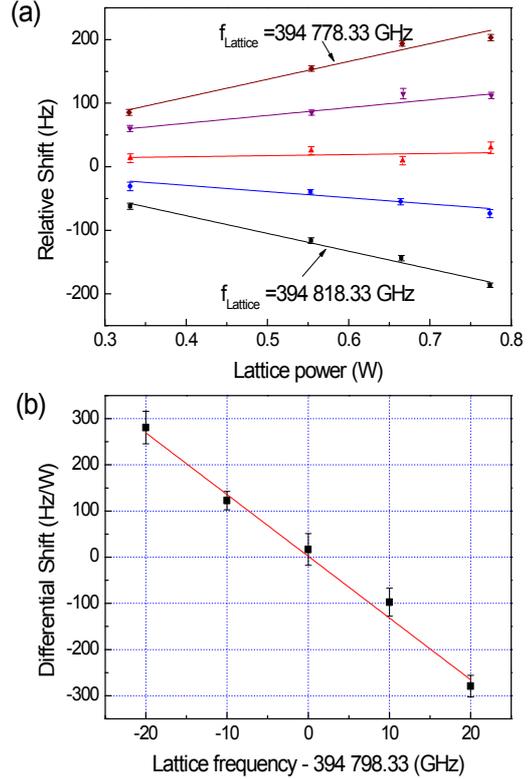

**Figure 6.** (a) Frequency shift of the clock transition depending on the lattice laser power at five different frequencies of the lattice laser evenly spaced by 10 GHz around the center frequency of 394 798.33 GHz. (b) The differential shift at five frequencies of the lattice laser; each point was obtained by the slope from the linear fit at a given frequency of the lattice laser.

To evaluate the light shift due to the lattice laser, we first determined the magic frequency at which the ac Stark shift of the clock transition due to the lattice laser vanishes. As shown in Fig. 6(a), the frequency shift of the clock transition was measured by variation of the lattice laser power from 0.3 W to 0.8 W at five different frequencies evenly spaced by 10 GHz around the center frequency of 394 798.33 GHz. Each data point in Fig. 6(a) was obtained from the average value of 40 measurements of the clock transition. Differential shifts for various lattice laser frequencies were obtained (Fig. 6(a)) and fit to find the magic frequency (Fig. 6 (b)) of 394 798.48 (79) GHz, which agrees well with previous reports [12, 13]. We performed three separate absolute frequency measurements on August 30th (MJD 55803), September 1st (MJD 55805), and October 14th (MJD 55848). We fixed the lattice laser frequency for all three measurements at 394 798.33 GHz, which was -0.15 GHz away from the determined magic frequency. The lattice beam power was at 680 mW for the first and second measurements and 540 mW for the third one. The frequency shift was 1.4 Hz with an uncertainty of 7.3 Hz and 1.1 Hz with an uncertainty of 5.8 Hz for each of the beam power used as shown in the uncertainty budget in Table 1.

Since hyperpolarizability shift could not be measured within our measurement uncertainty, we estimated the shift and uncertainty from the results obtained by other group with better precision [9, 13], in which the frequency shift was measured to be 0.17(4) Hz at 500 $E_r$ depth. Considering the reported shift and the shallower lattice potential depth of 420 $E_r$ (MJD 55803, 55805) and 340 $E_r$ (MJD 55848) used in our experiments, we estimated the associated shift to be 0.12(4) Hz and 0.08(3) Hz for each measurement day with different lattice beam power used as shown in Table 1.

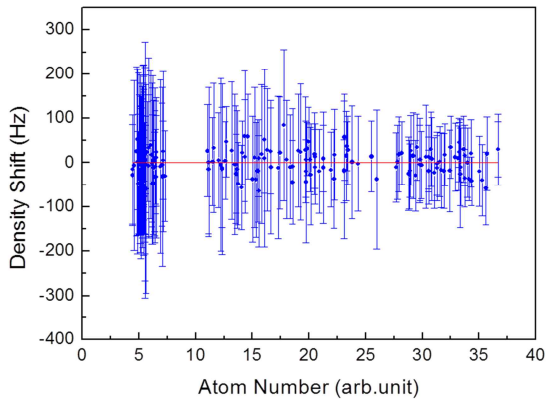

**Figure 7.** Measurement of collision shift of the clock transition. Red solid line is a linear fit to the data points.

The vector-polarizability-induced shift disappears with linearly polarized light [11]. Although a small amount of the polarization ellipticity can take place due to the birefringence of mirrors and vacuum windows, vector shift can be suppressed to less than $10^{-18}$ by the use of linear polarization, cancellation due to the averaging process of two π-transitions, and experimental geometry [13]. Therefore, we did not correct the vector shift in Table 1.

The first-order Zeeman shift was also cancelled by taking the average frequency of the two π-transitions in Fig. 5. However, the second-order Zeeman shift depends on the square of the magnetic field intensity, and it cannot be cancelled in this way. Using the value of the bias B-field of 430 μT measured from the Zeeman splitting and the second-order Zeeman coefficient of -7(1) Hz/mT$^2$ [13], the frequency shift due to the second-order Zeeman effect was estimated to be -1.3(2) Hz.

The blackbody radiation shift was evaluated by measurement of the temperature and the fractional solid angles (FSA) of the components surrounding the atoms in the lattice trap as discussed in Ref. [33]. Considering the stainless steel vacuum chamber (36~43°C, FSA=0.9), the view-port windows (26~29°C, FSA=0.1), and the furnace for the atomic beam (452(5)°C, FSA=0.001), the frequency shift due to the blackbody radiation was estimated as shown in Table 1.

The gravitational shift was evaluated using the height of the lattice trap (92.0 m) from the geoid, which was measured by a GPS antenna. The frequency shift to the gravity was estimated to be 5.2(1) Hz.

The collision shift should be included in the uncertainty budget, because approximately $10^5$ atoms were captured in several hundred lattice sites, and the estimated density was higher than $10^{11}/cm^3$. Since the atoms in a lattice were not spin-polarized in our measurements, s-wave collision between the ground state atoms was not prevented. A recent result [14] has shown that the collision shift of un-polarized atoms is smaller than that of spin-polarized ones, implying that competing interactions have the opposite sign as those in the polarized case. However, the excitation fraction was not fixed in our experiment, since the clock laser frequency was not locked at a specified excitation fraction but was scanned across the resonance peak to get the clock transition spectrum. Therefore, only the averaged shift for different excitation fractions could be measured. To experimentally determine the collision shift, we varied atom numbers by interleaving the loading time of the blue-MOT from 80 to 680 ms. The atom number (with an arbitrary unit) was measured by the use of integrated fluorescence signals (*A, B* in Fig. 4) of the 399-nm probe light, which were described in section 2.2. As shown in Fig. 7, the linear least-square-fit result gives the slope of 0.04 with the uncertainty of 0.20, which means that the density shift could not be clearly observed within our frequency measurement uncertainty. Under the typical experimental condition of our frequency measurement at an atom number of 14(3) in this arbitrary unit, the collision shift was estimated to be 0.6 Hz with the uncertainty of 2.7 Hz.

The light shift due to the clock laser can be estimated from the previous results [10]. If we adopt the reported measured coefficient and the clock laser power of 1.6 mW/cm$^2$ used in this work, the shift is negligible (less than 30 mHz). The other systematic effects, such as the residual Doppler shift and the AOM phase chirp, are estimated to be negligibly small within our frequency measurement uncertainty. The systematic shifts and associated uncertainties are summarized in Table 1 for each measurement day.

### 4. Absolute frequency determination

To measure the absolute frequency of the clock transition, we measured the frequency of the clock laser by the use of an optical frequency comb referenced to an H-maser. The frequency offset of the H-maser from TT and the frequency shifts due to the systematic effects, which were discussed in section 3, were compensated.

As already noted in section 3, 592 measurements were taken on three separate days over a 45 day period (241 measurements on MJD 55803, 187 measurements on MJD 55805, and 164 measurements on MJD 55848). The measurement result on MJD 55803 is shown in Fig. 8(a) after the correction of the systematic shifts and the H-maser offset from TT. The histogram of the frequency measurements is fit by a Gaussian distribution as in Fig. 8(b). Fig. 8(c) shows the statistical uncertainty (total deviation) versus the number of measurements (randomized). The uncertainty averages down as N$^{-0.50(2)}$. The corresponding statistical measurement uncertainties for each measurement were 1.8 Hz, 2.2 Hz, and 1.7 Hz, respectively.

The H-maser frequency offset from TAI for a 5-day interval was calculated from the internal time comparison data between H-maser and UTC(KRIS), and the time difference between UTC(KRIS) and UTC published in Circular T. Since the uncertainty due to the internal time comparison was negligible, the uncertainty was dominated by the link uncertainty for the time difference between UTC(KRIS) and UTC. The link uncertainty was calculated based on the recommendation from the Working Group on Primary Frequency Standards [34, 35] as 0.7 Hz for 5-day interval.

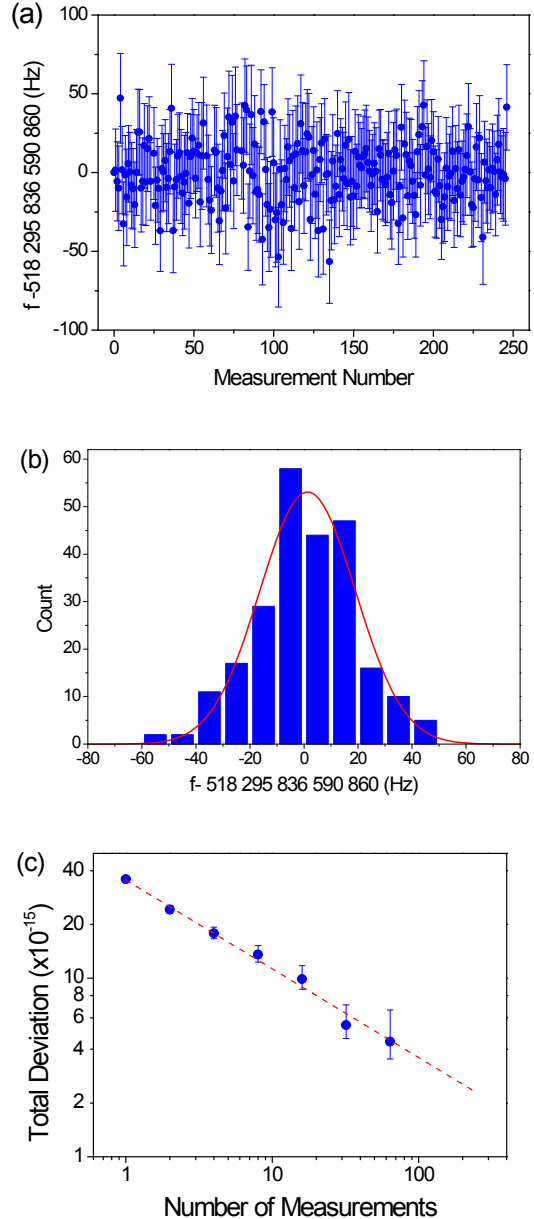

**Figure 8.** (a) 241 measurement results of the absolute frequency of the clock transition on August 30$^{th}$(MJD 55803) after correcting the frequency shift by the systematic effects and the H-maser offset from TT. (b) Histogram of the frequency measurements in (a) and its Gaussian fit. (c) The statistical uncertainty (total deviation) versus the number of measurements (randomized). The uncertainty averages down as N$^{-0.50(2)}$.

**Table 1.** Uncertainty budget for the absolute frequency measurement of the $^{171}$Yb optical lattice clock for three different measurement days – MJD 55803, 55805, and 55848.

| Effect | MJD 55803 Shift (Hz) | MJD 55803 Uncertainty (Hz) | MJD 55805 Shift (Hz) | MJD 55805 Uncertainty (Hz) | MJD 55848 Shift (Hz) | MJD 55848 Uncertainty (Hz) |
|---|---|---|---|---|---|---|
| Linear lattice ac Stark shift | 1.4 | 7.3 | 1.4 | 7.3 | 1.1 | 5.8 |
| Hyperpolarizability | 0.12 | 0.04 | 0.12 | 0.04 | 0.08 | 0.03 |
| Second order Zeeman | -1.3 | 0.2 | -1.3 | 0.2 | -1.3 | 0.2 |
| Gravitational shift | 5.2 | 0.1 | 5.2 | 0.1 | 5.2 | 0.1 |
| Blackbody radiation shift | -1.6 | 0.2 | -1.6 | 0.3 | -1.5 | 0.2 |
| Collisional shift | 0.6 | 2.7 | 0.6 | 2.7 | 0.6 | 2.7 |
| **Yb** | **4.4** | **7.8** | **4.4** | **7.8** | **4.2** | **6.4** |
| Yb-Maser (statistical) | - | 1.8 | - | 2.2 | - | 1.7 |
| Yb-Maser(deadtime)[a] | - | 0.9 | - | 1.0 | - | 1.5 |
| Maser-TAI(5 day)[b] | 10.6 | 0.7 | 10.9 | 0.7 | 7.1 | 0.7 |
| TAI-TT[b] | 2.7 | 0.2 | 2.6 | 0.2 | 2.4 | 0.2 |
| **Total** | **17.7** | **8.1** | **17.9** | **8.2** | **13.7** | **6.8** |

a) Uncertainty due to the difference between the measurement time and frequency calculation period of 5 days in Circular-T.
b) Calculated from Circular-T No. 284, 285, and 286 respectively.

Since the measurement times were 36909 s, 28613 s, and 8915 s for each of the measurement days, and the frequency offset of the H-maser from TAI was calculated for a 5-day interval, we should estimate the dead time uncertainty. The noise characteristics, $\sigma_y(\tau)$, of the H-maser were estimated by three noise types, which are $8.0\times10^{-16}\tau^{-1/2}$ for white FM, $6.0\times10^{-16}$ for flicker FM, and $4.0\times10^{-16}\tau^{1/2}$ for random-walk FM, where $\tau$ is expressed in days. The dead time uncertainty was estimated considering the stability of the H-maser as 0.9 Hz, 1.0 Hz, and 1.5 Hz for each day [36].

The TAI frequency offset from TT was calculated from the duration of the TAI scale interval of Circular T covering each measurement day (Circular T No. 284, 285, and 286) as $-5.2(3)\times10^{-15}$, $-5.0(3)\times10^{-15}$, and $-4.6(4)\times10^{-15}$, which resulted in shifts of 2.7(2) Hz, 2.6(2) Hz, and 2.4(2) Hz.

The absolute frequency measurement results for all days are shown in Fig. 9. We compensated the corresponding systematic frequency shifts, and the error bar indicates the total uncertainty. From the weighted mean of the three measurements with weight proportional to the inverse of the total uncertainty squared, the absolute frequency of the clock transition was determined to be 518 295 836 590 863.5 Hz. To estimate the final total uncertainty, we first calculated the statistical measurement uncertainty $u_s$ as $1/u_s^2 = \sum 1/u_i^2$, where $u_i$ is the statistical measurement uncertainty for each day, resulting in 1.1 Hz. For systematic, dead time, link, and TAI offset uncertainties, we assigned the maximum values for each effect from all three measurement days. The final total uncertainty was estimated to be 8.1 Hz ($1.5\times10^{-14}$). The solid red line shows the weighted mean value, and the dashed line shows the total uncertainty. This result is compared with the two previous measurements by other groups in Fig. 10, which shows good agreement within the uncertainty of this work.

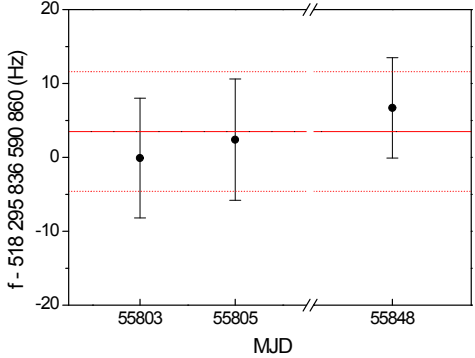

**Figure 9.** The three absolute frequency measurement. The determined absolute frequency and uncertainty from the weighted mean of the results are indicated by the red solid line and dotted line, respectively.

## 5. Conclusion

We measured the absolute frequency of the $^1S_0(F = 1/2) - {}^3P_0(F = 1/2)$ transition of $^{171}$Yb atoms confined in a 1-D optical lattice. The determined frequency was 518 295 836 590 863.5 Hz with an uncertainty of 8.1 Hz ($1.5 \times 10^{-14}$). As the third independent measurement, our result is in good agreement with the results of previous measurements of other laboratories within the uncertainty of this work.

To reduce the current frequency measurement uncertainty, which is limited mainly by the clock laser instability, an improved clock laser system with better stability will be employed in the near future. The support point of the cavity, the control loop bandwidth in the linewidth reduction, and the stabilized temperature of the cavity will be optimized. Also, a Cs fountain clock is under development at KRISS, with which the uncertainty due to the frequency transfer from TT can be eliminated. We expect that a more precise frequency measurement is within our reach after these improvements and direct comparison of optical lattice clocks with other laboratories become possible.

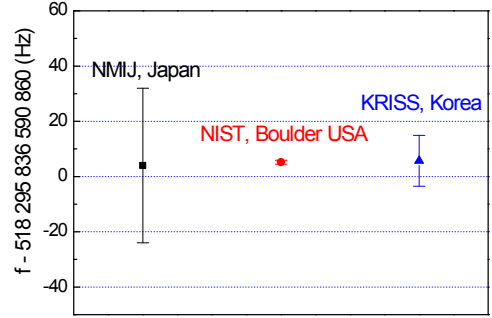

**Figure 10.** Measurement results of the absolute frequency of the $^{171}$Yb lattice clock transition by different institutes: NMIJ [12], NIST [13], and KRISS (this work).

*Note added.*—While submitting this manuscript, we became aware of another recent absolute frequency measurement of the $^{171}$Yb clock transition by NMIJ [37] with improved uncertainty, which agrees with three values shown in Fig. 9.

## Acknowledgements

We gratefully acknowledge the experimental assistance of Masami Yasuda during his stay at KRISS and helpful discussions with Feng Lei Hong from NMIJ. This work was supported by Korea Research Institute of Standards and Science under the project of 'Establishment of National Physical Measurement Standards and Improvements of Calibration/Measurement Capability', grant 12011002.